\documentclass{ws-ijmpa}

\begin{document}

\markboth{H. Neuberger}{Lattice Chirality and Large N}
%
\catchline{}{}{}{}{}
%

\title{Lattice Chirality and its uses at large $N_c$}

\author{Rajamani Narayanan}

\address{Florida International University, Department of Physics,
Miami, FL 33199. \\ email: rajamani.narayanan@fiu.edu}

\author{Herbert Neuberger\footnote{Speaker at the International Conference on QCD 
and Hadronic Physics, June 16-20, 2005, Beijing.}}

\address{Rutgers University, Department of Physics
and Astronomy,
Piscataway, NJ 08855. \\ email: neuberg@physics.rutgers.edu}

\maketitle

\begin{abstract}
A brief overview of the authors' work on lattice chirality
and its application to the numerical study of planar QCD
is presented.
\end{abstract}

\section{Lattice Chirality} 

Path integrals are formal objects. Sometimes one says that the 
integration measure is ill defined, 
to indicate that we do know the classical action, but, 
as we know quite well, classical Lagrangians simply are not 
guaranteed to produce unique quantum mechanical systems. 
Separating the problem into one of measure versus one of classical action 
is more a psychological step than a concrete mathematical statement; 
the fundamental construct is the quantum theory. The path integral 
needs to be defined as a whole, and the separation into a measure
term and the rest is ambiguous.

In the mid 80's mathematical physicists provided a 
concrete view of the path integral over Weyl fermions 
in Euclidean space-times with non-abelian gauge backgrounds: 
The result of ``performing'' the fermion integral is not a function 
of the background, but, rather, a U(1) bundle defined over 
the space of all backgrounds. Anomalies indicate when the bundle can be 
trivialized. The bundle itself is defined over the space of gauge orbits 
and is therefore tautologically gauge invariant. Anomalies affect the choice 
of sections to the bundle: to do the remaining functional integrations one 
needs to pick a section.

Most lattice field theorists ignored these developments because anomalies 
were discovered and understood much earlier and without using anything close 
to the latter insight. Moreover, the new insight was specific to the Euclidean 
version of the theory.
We know now that this was a mistake, but that was not obvious at the time. 
What one failed to appreciate was that the lattice made 
gauge invariance exact, 
and, in the Euclidean action formalism, there was no 
escape from a more radical 
change of view about what a lattice fermion integral 
really meant: For example, an 
integral over an odd number of fermion fields had to 
be permitted to give a non zero answer!

The successful resolution of the problem of lattice 
chirality is rooted in a paper by
Callan and Harvey who showed how anomalies on embedded 
defects are realized by current inflows. 
Later, David B. Kaplan showed that the Callan Harvey 
effect works on the lattice with Wilson fermions.
Almost simultaneously with Kaplan, and completely independently,  
Frolov and Slavnov came up with the idea that in order 
to regulate a chiral gauge theory one needs a strictly 
infinite number of extra regulator fields. An old paper by
Ginsparg and Wilson presciently observed that 
global chiral symmetries could be deformed on the lattice 
without losing any of their physical consequences 
and the modern development realized that observation
in the case of QCD. 

The objective is to define a chiral determinant 
line bundle on the lattice, over the space of all 
gauge orbits. The space of gauge orbits needs to have some subsets excised to 
permit the existence of nontrivial bundles. The infinite number of four 
dimensional fermion fields in the Callan and 
Harvey, David B. Kaplan and Frolov 
and Slavnov setups make the path integrals 
ambiguous to the extent that they do 
not always produce a gauge invariant function. 
However,  they can always be used 
to gauge invariantly construct a line bundle.

In the vector like case the product of the bundle and its conjugate produces a 
function which turns out to be the determinant of a matrix that obeys the 
Ginsparg Wilson version of global lattice chirality. That matrix is called 
the ``overlap'' Dirac operator.

A U(1) bundle is a smooth collection of one dimensional 
complex vector spaces. A direct way to define 
one is via a set of linear equations, 
in which the parameters enter smoothly, and 
which have a one dimensional space of 
solutions, that is one less equation than unknowns. 
Such are the equations that define 
eigenvectors of a matrix. In our context the 
matrix is the Wilson Dirac operator with mass set to 
negative unity and r-parameter set to one, 
$H_W$. Close enough to continuum, $H_W^2$ is 
bounded away from zero and it is meaningful 
to define a projector on the subspace 
spanned by eigenvectors of $H_W$, $v_i$, with 
positive eigenvalues. For gauge fields that 
can be deformed to zero, the number of positive 
and negative eigenvalues of $H_W$ are 
equal. (If there is a difference, it gives 
the topological charge of the background.) 
Let us assume that the topological charge is 
zero. Let $w_j$ be the eigenvectors of $H_W$ with 
positive eigenvalues for a positive Wilson mass. 
The line bundle is given by:
\begin{eqnarray*}
\det_{j,i} <w_j|v_i>
\end{eqnarray*}

The bundle is defined only in terms of projectors
on the positive eigenspaces, $P_+$ and $P_-$, 
and therefore it is not surprising that 
the {\it function} giving the 
absolute value of the bundle can be 
expressed in terms 
of these projectors solely, or, equivalently, in terms of 
$\epsilon_\pm=1-2P_\pm$: $
|\det_{j,i} <w_j|v_i>|^2= \det\frac{1+\epsilon_+ \epsilon_-}{2}$.

The bundle can be nontrivial:
One has to excise orbits for which $H_W$ has zero modes (the projectors are 
ill defined for those). 
One cannot deform smoothly $P_-$ by taking the Wilson mass to an extremal 
value to simplify $H_W$. However, $P_+$ can be deformed and nothing is lost by 
replacing $\epsilon_+$ by $\gamma_5$. $P_-$ must remain nontrivial, 
with $\epsilon_- = sign(H_W)$. The unitary 
matrix $V=\gamma_5 \epsilon_-$ becomes the central object.

The Nielsen Ninomiya theorem is avoided because one can use slightly 
different propagators for the sea and valence quarks. The sea quarks have a 
quadratic action with kernel usually denoted by $D_o$, the subscript standing 
for the letter $o$ from the word overlap ($<w|v>$). The valence quarks have 
a propagator that anticommutes with $\gamma_5$. It is highly nontrivial that
one can use different propagators for the fermions 
in this manner. The proof rests
on an argument in which one introduces an auxiliary field and declares that
all Green's functions of interest are correlations of one particular
linear combination of the fermion fields and the auxiliary field. The 
multiplicative contribution
of the auxiliary field to the 
total fermion determinant is unity. In the continuum limit
the auxiliary field is infinitely massive and there is no remnant of using the
linear combination of fields as interpolating fields 
for the external quark lines.

The overlap Dirac operator $~D_0=\frac{1+V}{2}$ is built out of the
unitary matrix $V$, which is not sparse. 
The propagator on fermion lines connected to external
quarks (valence quarks) is given by the gauge covariant matrix $A^{-1}$ which
anticommutes with $\gamma_5$. $A^{-1}=\frac{1-V}{1+V}$.

The net practical result is that it becomes possible for the first time
to establish $S\chi SB$ breaking at finite ultraviolet cutoff, independently 
of the approach to the continuum limit and any 
explicit chiral symmetry breaking.

\section{Uses at large $N_c$}

Eguchi Kawai reduction gave hope 
that one would be able to solve QCD 
at infinite $N_c$, still by numerical means, 
but faster than true QCD. The main 
stumbling block were lattice chirality 
and topology. These problems are now 
out of the way.

Consider YM at $N_c=\infty$ on a Euclidean torus of size $l^4$ 
(this space has a hyper-cubic 
symmetry group, preserved 
by typical lattice regularizations). There exists a physical length, 
$l_c$, such that
traces of Wilson loops associated with loops $C$, $W( C )$, are exactly $l$ 
independent at $N_c=\infty$  for $l>l_c$.
Hence, the finite $l$  looks infinite. 
For $l>l_c$, the entire extra symmetry group associated 
with an $SU(N_c )$ gauge theory on a
four-torus,  $Z^4(N_c )$, is unbroken and so is the hyper-cubic symmetry group.

On the continuum torus, momentum space is non-compact, 
but discrete, with gaps $2\pi/l$.
The eigenangles of Polyakov loops, $\theta_\mu^i$, 
contribute distinguishable 
momenta $|\Delta p_\mu^i| < 2\pi k_\mu / (N_c l)$, where $|k_\mu|<N_c /2$.
So long as the $\theta_\mu^i$ are uniformly distributed, 
the $|\Delta p_\mu^i |$ exactly fill the $2\pi /l$ gaps.
The $\theta_\mu^i$ will be uniformly distributed if the 
$Z^4(N_c )$ remains unbroken.

As $l$ is lowered to $l_c$, one of the four directions 
is picked randomly and the 
associated $Z(N_c )$ gets spontaneously broken. The transition is 1-st order.
The eigenvalue spectrum of the Polyakov loop opens a gap at a random location.
The infinite $N_c$ system suddenly ``discovers'' that one 
of the four directions is 
finite, but it still ``believes'' that the others are infinite.
This has a precise meaning: the $N_c=\infty$ system represents finite 
temperature planar QCD, $T>l_c^{-1}$.

\begin{figure}
\centerline{\psfig{file=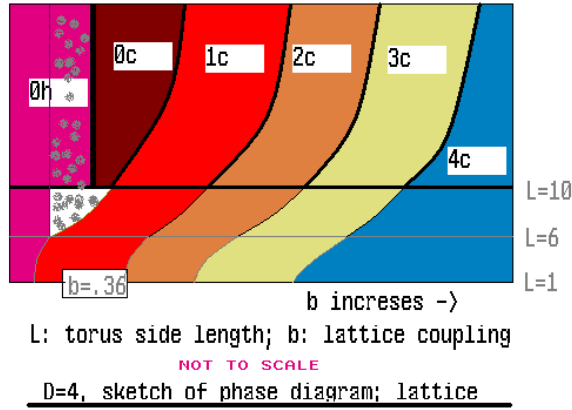,width=8cm}}
\caption{A schematic description of the large $N$ phases 
of planar QCD on a four torus.}
\end{figure}

Figure 1 shows the various phases on the 4 dimensional lattice. 0h only exists on
the lattice and disappears in the continuum. 
There are five different Xc phases, with X ranging from 0 to 4.
As X increases beyond 0, sequentially, a Polyakov loop in one additional direction opens a gap, 
as sketched in Figure 2. The 0h to 0c transition is strongly first order and
0c extends into the 0h phase by metastability. The
boundaries separating Xc phases satisfy asymptotic scaling and
define separate critical sizes (exponential in the inverse gauge
coupling). In all Xc phases, $\|W(1,1)-1\| \le e$ where $W(1,1)$ is
the smallest Wilson loop and $e < 1$ approaches zero as $N_c\to\infty$; this
produces an unambiguous definition of topology. 
There is no 0c phase for $L < 5$ and
in particular for $L=1$. 
That there is a 0c phase for $ L > 4$
was missed in past work on reduction.

\begin{figure}
\centerline{\psfig{file=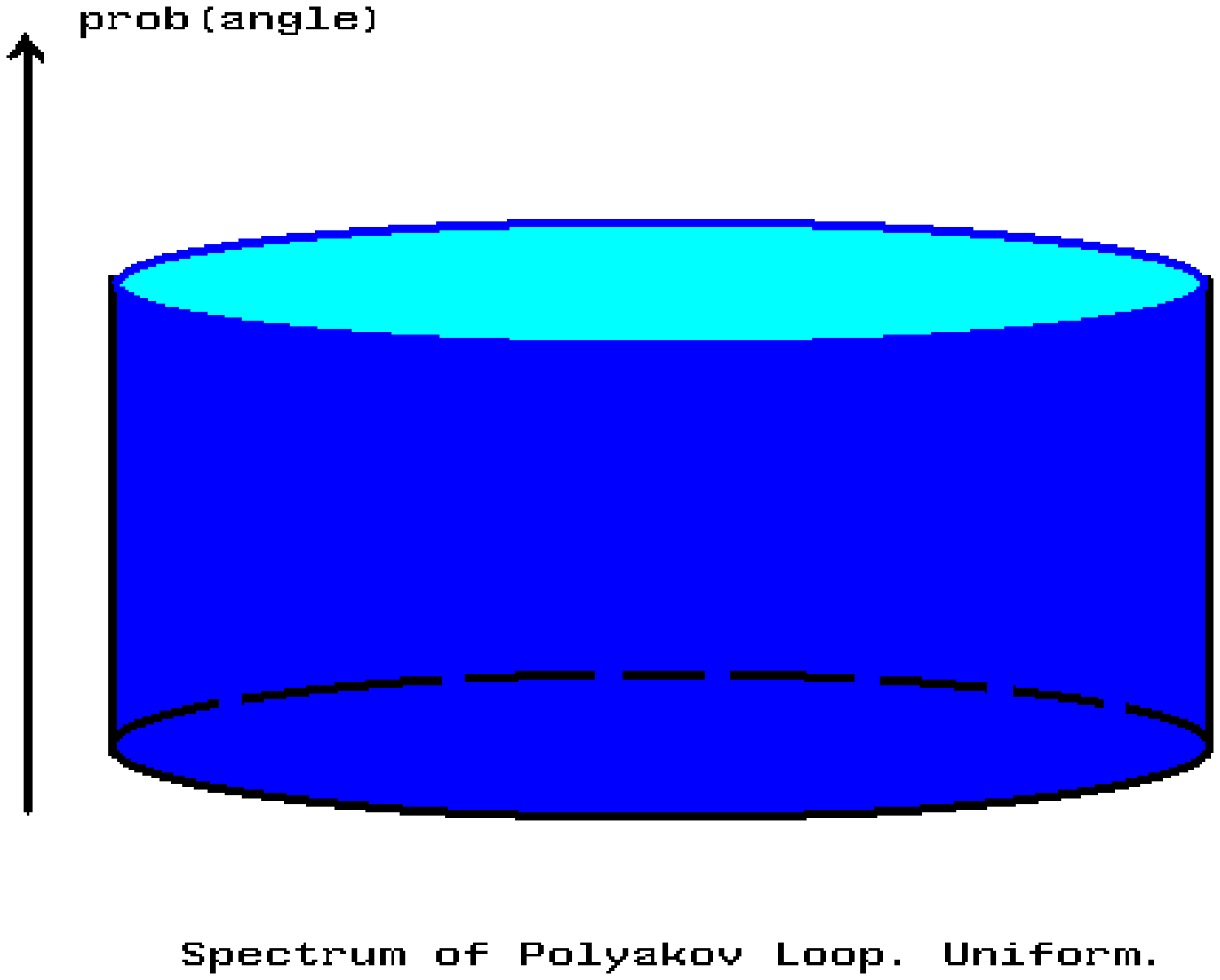,width=6cm}}
\centerline{\psfig{file=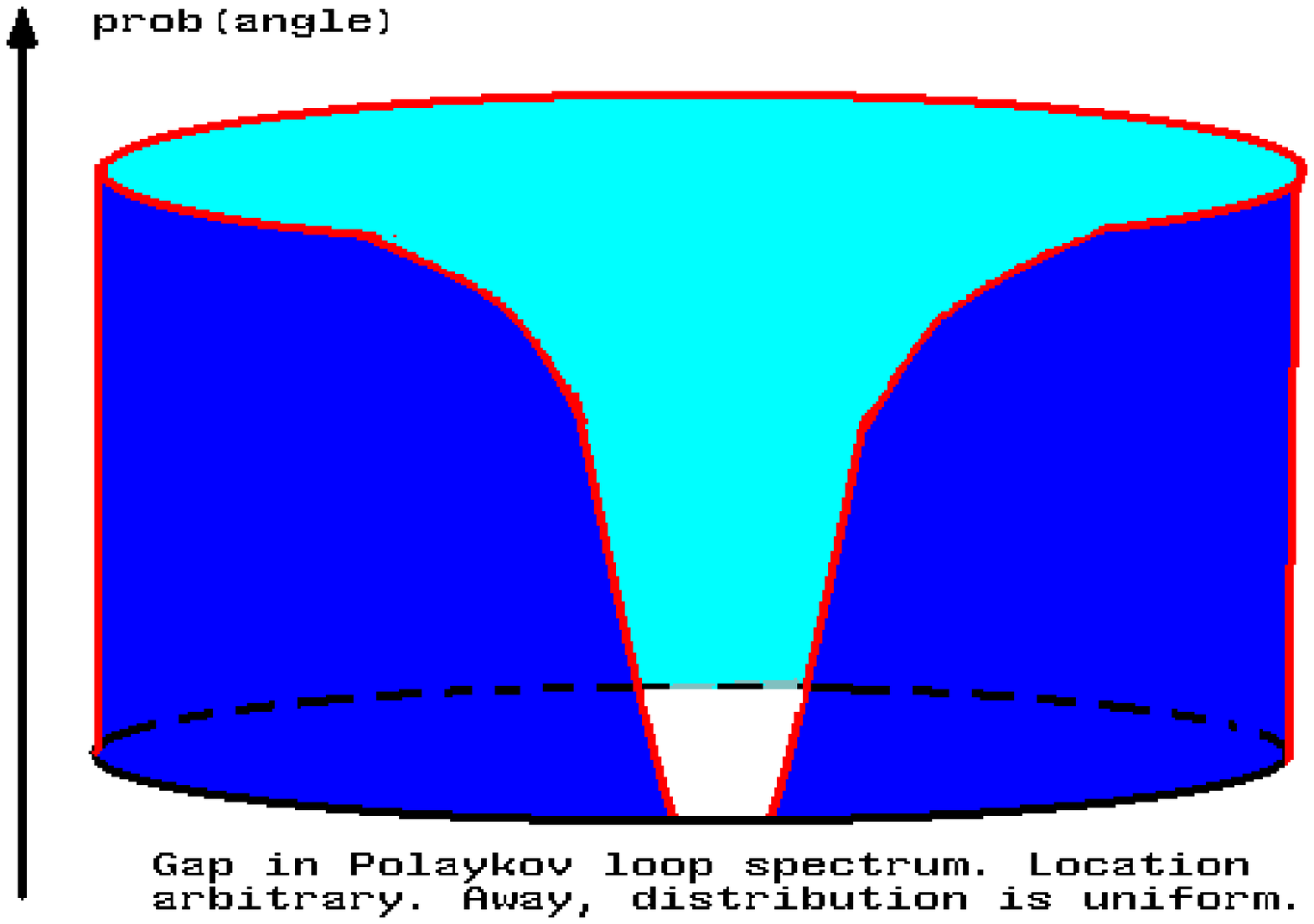,width=5.5cm}}
\caption{A schematic illustration of the breaking of $Z(N_c)$ in some direction. 
}
\end{figure}

\begin{figure}
\centerline{\psfig{file=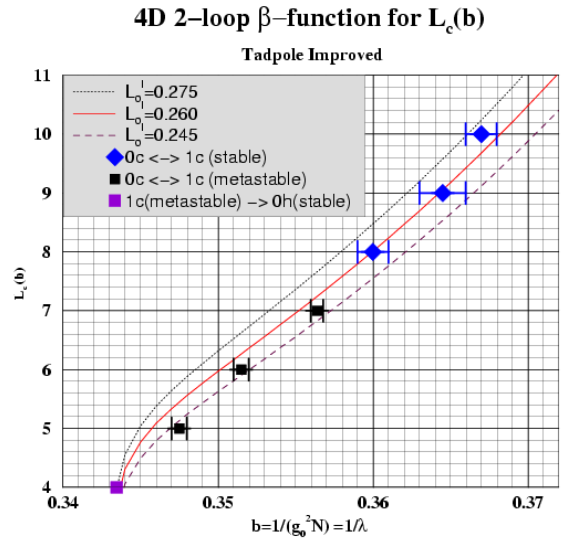,width=8cm}}
\caption{A plot of the critical lattice size as a function of lattice coupling;
obviously the size is an integer and the Monte Carlo simulation was used to
find the corresponding gauge coupling $b$.}
\end{figure}

In Figure 3 the Monte Carlo data for the 0c to 1c transition is
plotted and compared to tadpole improved asymptotic scaling, given by:
$$
b_I=b\frac{b^2-0.58960b+0.08467}{b^2-0.50227b+0.05479}
;\ \ \ \ L_c(b)=(0.260\pm 0.015)
\left (\frac{11}{48\pi^2b_I} \right )^{\frac{51}{121}} 
e^{\frac{24\pi^2b_I}{11}}
$$

The massless Dirac operator takes the form
$A=\pmatrix{0&C\cr-C^\dagger&0}$
in the chiral basis with $C$ being a complex matrix.
The size of $C$ is  $n=2^{d/2}L^d N_c$, 
where $d$ (even) is the dimension.
Shuryak and Verbaarschot 
made C random, with enhanced symmetry:
$p(C)d^{2n^2} C \propto e^{-\kappa^2 n {\rm Tr} (C^\dagger C)}
d^{2n^2} C~.$
The model correctly captures the distribution of the 
small eigenvalues of $C^\dagger C$, up
to one scale determined by the chiral condensate.

It was determined by numerical simulation that the distribution of the two 
lowest eigenvalues of $-A^2$ indeed is given by 
RMT with a nonzero $\kappa$, at finite 
$L$ and infinite $N_c$. $\kappa$ is independent of $L$. 
This establishes spontaneous chiral 
symmetry breaking and large $N_c$ reduction for the chiral condensate:
$\frac{1}{N_c} \langle\bar\psi\psi\rangle=\kappa$.

In the two dimensional 't Hooft model (planar QCD) 
the meson mass, $m^2 (m_q)$, contains enough 
information to almost reconstruct the full 't Hooft's meson equation.
That equation can be ``localized'' by adding extra degrees of freedom. 
Broadly speaking, an extra dimension also localizes meson 
wave equations in AdS duals. 
Assume that four dimensional $N_c=\infty$ QCD is described by a dual 
with a warp factor, one scalar 
real function of one scalar real variable as conjectured by Polyakov.
Could a meson probe determine the unknown warp function by something akin to 
inverse scattering, given $m^2(m_q)$ in the planar limit ?
We analyzed the MC data using a dimensionless and renormalization group-invariant parameterization:
\begin{eqnarray*}
&\Delta=\frac{1}{2} [\sqrt{m_\pi^2 L_c^2(b) + \Lambda_\pi^2}-\Lambda_\pi];
~~\frac{1}{4} m_\pi^2L_c^2(b)=\Delta(\Delta+\Lambda_\pi )\\
&\Delta= m_o\Sigma(b)L_c^4(b) +\frac{1}{\Lambda_q} 
m_o^2\Sigma^2(b)L_c^8(b) +....\\
\end{eqnarray*}
Fits to the data produce: $\Lambda_\pi=6.91~~ \Lambda_q=1.03$. 
This also allows the extraction of the pion
decay constant at infinite $N_c$ and
we see that it has a large 
finite $\frac{1}{N_c}$ correction at $N_c=3$:
$f_\pi (QCD) \approx \sqrt{3}\frac{f_\pi (N_c=\infty)}
{\sqrt{N_c}}=123~MeV~\ne~86~MeV$.

At ${N=\infty}$ $m^2(m_q)$ is well described by a parabola 
for all $m_q$, well beyond the 
range of chiral perturbation theory. 
The parameter $\Delta$ enters in a way familiar from mass formulae
obtained using the supergravity dual to the planar limit of certain
field theories, but no longer is defined in terms of a dimension. This
and the observations about the 't Hooft model lead to the question: 
Does the parameterization of $m^2$ in terms of a $\Delta$ 
indicate that the meson 
wave equation can be approximately localized via the 
introduction of extra degrees of freedom ?

\section{Acknowledgments}

This research was partially supported  
by the DOE under grant number 
DE-FG02-01ER41165 at Rutgers University
and by the NSF under grant number PHY-030065 at FIU.
Herbert Neuberger is grateful to the conference organizers for
the invitation.

\end{document}